\documentclass[aps, singlecolumn, showpacs]{revtex4-2}
\usepackage{amsmath}
\begin{document}
\title {Harmonic Spinors and $Z_2$ Vortex}

\author{S. C. Tiwari \\
Department of Physics, Institute of Science, Banaras Hindu University, Varanasi 221005, and \\ Institute of Natural Philosophy \\
Varanasi India\\ Email address: $vns\_sctiwari@yahoo.com$ \\}
\begin{abstract}
Hodge theorem and harmonic spinors are studied in a physics-oriented approach in the present paper. New mathematical results on the harmonic spinors are as follows.  Harmonic spinors defined by partial differential operators could be of two types: trivial without topological defects, and having nontrivial topological structures, for example, phase singularities or phase vortices. There could exist a nontrivial harmonic vector field associated with nontrivial harmonic spinor, for example, ${\bf v}_{vortex}$ associated with Weyl 2-spinor. The $Z_2$-vortex is re-visited in the perspective of harmonic spinors leading to a remarkable result that the gauge potential is exactly the same as the nontrivial harmonic vector field associated with the 2-spinor. It is proposed that a discrete symmetry group $SL(2, Z)$ has a role in connection with the continuous group $SU(2)$ similar to the discrete group $Z_2$ in $U(1)$.


\end{abstract}

\maketitle

\section{\bf Introduction}

The imaginary unit $i=\sqrt{-1}$ and the spinors continue to be the objects of interest, intrigue and investigations both for the mathematicians and physicists. Historically, Cartan in 1913 introduced spinor as a two-valued representation of orthogonal group; the name spinor was  later given by Ehrenfest  \cite{1}. Penrose \cite{1} points out that the related ideas of Hamilton's quaternion in 1837 and the Clifford algebras in 1878 predate Cartan spinors. Clifford algebras could be seen as the generalization of quaternions and Grassman algebras of 1825. Hestenes remarks \cite{2} that though the geometric significance of the algebras was common in the works of Hamilton and Grassman, a "harmonious whole" of the two was not made. Hestenes proposes geometric algebra for the suggested synthesis. An analogue of Hodge theorem for spinors was studied in PhD thesis by Hitchin \cite{3}. There is a vast literature on the geometry and the algebras of the spinors. However, relatively less attention is paid to this topic in physics. 

The aim of the present paper is to unravel new insights on the topology of spinors. Besides the idea of harmonic spinors, there is an interesting question on the parallel spinors and holonomy groups. Salamon \cite{4} notes that parallel spinors can distinguish holonomy for symmetric and non-symmetric cases. Cohomology group and Hodge decomposition play important role in understanding the topological characteristics of a manifold using exterior differential forms. To appreciate harmonic spinors and holonomy groups \cite{3,4}, as well as to investigate the significance of harmonic vector field in connection with the harmonic spinors , a brief introductory discussion on de Rham cohomology and differential forms is presented in the next section following a nice textbook treatment \cite{5}. Essential rudiments can also be found in \cite{6}. 

Theory of harmonic spinors is developed in a physics-oriented approach in Section 3. It is noted that the Dirac 4-spinor in flat space-time is a harmonic spinor by definition. A 2-spinor satisfying Weyl equation is considered to derive nontrivial harmonic spinor solutions possessing phase singularity. The phase singularity is interpreted as a singular vortex with nontrivial topology. An important result is obtained that a nontrivial harmonic spinor is associated with a harmonic vector field. In Section 4, the so-called $Z_2$-vortex \cite{7} is studied in the perspective of the harmonic spinors. A remarkable result is obtained on the nature of the gauge potential in terms of the harmonic vector interpreted as a velocity field. In section 5, the relationship between continuous Lie groups and discrete groups is discussed based on the relation between $U(1)$ and $Z_2$ implied by the physics of $Z_2$ vortex. Concluding remarks constitute the last section.

\section{\bf de Rham period and Hodge theorem}

In order to simplify the discussion first we take an illustrative standard example \cite{5}. Let us consider a vector field 
\begin{equation} 
{\bf A}_h = -\frac{y}{r^2} \hat{i} + \frac{x}{r^2} \hat{j} + 0 \hat{k}
\end{equation}
In a truncated 3D space, origin removed, $R^3 -\{O\}$, both divergence and curl of ${\bf A}_h$ vanish, and the vector field satisfies the Laplace equation. Here $r^2 =x^2 +y^2$ and $\hat{i},\hat{j}, \hat{k} $ are unit vectors. It is a harmonic vector field, and the loop integral encircling the origin has a nonzero value equal to $2\pi$. The third dimension being redundant for the form (1) of the vector field, one may view this vector field in 2D space and re-write it as a differential 1-form  $d(\arctan \frac{y}{x})$. Now, in $R^2$ this form is not exact as  $\arctan \frac{y}{x}$ is not a differentiable function. However, it is differentiable on $R^2-\{O\}$. In the vector calculus, one has an important result that if the divergence and curl of a vector field are zero, the vector field satisfies the Laplace equation. If divergence is zero then the vector field may be defined as a curl of some vector field, and if the curl is zero, one can define the vector field as a gradient of a scalar field. 
A harmonic vector field, for example, ${\bf A}_h$ satisfies the Laplace equation. Another simple example of a harmonic vector field is
\begin{equation}
{\bf A} = y \hat{i}+ x \hat{j} +0 \hat{k}
\end{equation}
It is easily verified that this vector field is divergenceless  and curl-free, and can be written as a gradient of a scalar field ${\bf A}= {\bf \nabla}(xy)$. The field is nonsingular in whole domain of $R^3$ and there is no nontrivial topology. The form of the vector field (1) is of interest in singular fluid flows and current flows in physics.

A transition to the language of differential forms becomes intuitively understandable in the light of the preceding preliminary discussion on vector calculus. Scalar function $f$ is a 0-form, and its exterior derivative gives a 1-form $\alpha =df$. One gets following relations
\begin{equation}
ddf \equiv {\bf \nabla} \times ({\bf \nabla} f) =0
\end{equation}
\begin{equation}
dd\alpha \equiv {\bf \nabla} \cdot ({\bf \nabla} \times {\bf V})=0
\end{equation}

Formally, one has $d^2=0$. A totally antisymmetric covariant p-tensor field using the exterior or wedge product defines a p-form $\omega$.
The Hodge star operator defines a unique dual (n-p)-form $^*\omega$ on an oriented manifold $X^n$. The star operator is used to define a linear differential operator $\delta$ of degree $-1$. Note that a linear differential operator of degree $s$ is a mapping $T: A \rightarrow A$ such that $TT=0,~T: A^p \rightarrow A^{p+s}$. It can be proved that $\delta \delta =0$. For a vector field, we have $\delta \alpha \equiv -{\bf \nabla} \cdot {\bf V}$. The operator
\begin{equation}
\Delta = d\delta +\delta d
\end{equation}
is called the Laplacian. A differential form is harmonic if
\begin{equation}
\Delta \omega =0
\end{equation}

Integration of forms is embodied in generalized Stokes theorem
\begin{equation}
\int_c d\omega = \int_{\partial c} \omega
\end{equation}
The integration over chain $c$ of $d\omega$ is equal to the integration over the boundary $\partial c$ of $\omega$. The generalized Stokes theorem (7) shows that homology (the topology of the integration domains) and the cohomology (the topology of the differential forms) are dual to each other. The chain $c$ and the boundary $\partial c$ are coherently oriented, and the formula (7) is applicable for integrals of p-forms over p-chains.

The set of differential forms $\{ \omega^i\}$ and the set of chains $\{ c_i \} $ have a graded vector space
\begin{equation}
d\omega^i \subset \omega^{i+1} ~~ (d^2 =0)
\end{equation}
\begin{equation}
\partial c_i \subset c_{i+1} ~~ (\partial^2=0)
\end{equation}
The generalized Stokes theorem shows that $d^2=0 ~\Rightarrow ~\partial^2=0$.

The Poincare lemma states that $d(d\omega)=0$. For a p-form on manifold X, a (p-1)-form $\alpha$ such that $d\alpha =\omega$ implies $d\omega =0$. The form $\omega$ is called a closed form or a cocycle. The converse of the Poincare lemma, if $d\omega =0$ then $\omega=d\alpha$ may not be true globally. A form $\omega$ defined as $\omega =d\alpha$ is called an exact form or a coboundary. 

For the chains we define a cycle by $\partial c=0$ and a boundary by $c=\partial B$. The important results could be stated as follows
\begin{equation}
c=\partial B ~\Rightarrow ~\partial c=0
\end{equation}
but not necessarily
\begin{equation}
\partial c =0 ~ \Rightarrow ~ c=\partial B
\end{equation}
For the set of differential forms
\begin{equation}
\omega =d\alpha ~\Rightarrow ~ d\omega=0
\end{equation}
however, in general, it is not necessary that
\begin{equation}
d\omega =0 ~\Rightarrow ~ \omega =d \alpha
\end{equation}
The above results define homology and cohomology on X. A set of equivalence classes of cycles $Z_p$ of degree $p$ that differ by the boundaries $B_p$ i.e. $H_p(X) =Z_p/B_p$ defines homology. The de Rham cohomology $H^p(X)$ is defined as the set of equivalence classes of closed differential forms $Z^p$ modulo the set of exact forms $B^p$, i.e. $H^p=Z^p/B^p$. The de Rham theorem states the duality of the vector spaces $H_p$ and $H^p$ , and $b_p =b^p$. Here the dimension of $H_p(H^p)$ is called the Betti number $b_p(b^p)$.

In the book \cite{5} de Rham period is defined as the integral of a closed form over a cycle. One may state first and second de Rham theorems as follows. 

{\bf I:~} For a given set of periods $\{\nu_i\}$, there exists a closed p-form such that 
\begin{equation}
\nu_i = \oint_{c_i} \omega
\end{equation}
where $\{c_i\}$ is a set of independent cycles. The closed form is determined up to the addition of an exact form.

{\bf II:~} If all the periods of a closed form vanish then it is exact.

The reformulated de Rham theorem given in \cite{5} is reproduced: "There exists a closed form which has on $X^n$ arbitrarily preassigned periods of linearly independent homology classes. This closed form is determined up to the addition of an exact form." The corollary of this theorem is that a closed form is exact if and only if all its periods vanish. In physics literature, the concept of de Rham period is not widely used; Kiehn and Post are exceptions, see \cite{8}.

In general, a p-form on a compact manifold $X$ can be decomposed as a sum of closed, coclosed, and harmonic forms
\begin{equation}
\omega = d\alpha +\delta \beta +\gamma
\end{equation}
where the harmonic form $\gamma$ satisfies the Laplace equation
\begin{equation}
\Delta_p \gamma =0
\end{equation}
The Hodge theorem defines the cohomology class as follows. The space of harmonic p-forms on a compact manifold $X$ is isomorphic to the cohomology space $H^p(X)$. If $b_p$ is the p-th Betti number of $X$ then
\begin{equation}
b_p = dim ~ ker ~ \Delta_p
\end{equation}

It is noted in \cite{5} that the Hodge theorem relates analytical property of the differential operator $\Delta_p$ on $X$, the dimension of its kernel, and the topological property of $X$. The Atiyah-Singer index theorem is a generalization of Hodge theorem to an elliptic complex. An elliptic complex is defined by the elliptic operator
\begin{equation}
\Delta_p = D_p^* D_p +D_{p-1} D^*_{p-1}
\end{equation}
Note that the adjoint operator $D^*_p$ of $D_p=d$ is the operator $\delta$, and the above equation is just Eq.(5). Let $P(f) =0$ be an elliptical system of partial differential equations, then the analytical index is $index(P) = dim~ kernel(P) - dim~cokernel(P)$ and the theorem states that analytical index(P) = topological index(P).

\section{\bf Harmonic Spinors}

Does there exist an analogue of Hodge theorem for spinors? Formal similarity of Dirac operator with the differential operators $(D^*_p, D_p)$
implies that it is mathematically natural to investigate the dimension of null space, the harmonic spinors, and topological invariants for spinors \cite{3}. In general, the dimension of the space of harmonic spinors is not independent of the metric of the Riemannian manifold. However, Dirac operator is invariant under the conformal transformation of the metric. Even if the scalar curvature is zero, the homogeneous Riemannian manfolds have no harmonic spinors. Dirac operator $P=e_i \nabla_i$ is a covariant derivative along the direction $e_i$, and it is self-adjoint. The operator $P^2$ is the spinor Laplacian. The Dirac operator is an elliptic operator, and $ker~P=H$ is the finite dimensional space of harmonic spinors.

In a spin manifold with scalar curvature zero, every harmonic spinor is parallel, i. e. its covariant derivative vanishes. Recall the definition of holonomy group: parallel transport around a loop based at a point on the manifold $M^n$ defines the holonomy. Reduction of the holonomy group, for example, by a parallel vector field \cite{4}  leads to a product $R \times M^{n-1}$.  Salamon \cite{4} observes that parallel spinors can distinguish symmetric and non-symmetric spaces, and for a manifold with nonvanishing scalar curvature, there exist harmonic spinors but not parallel spinors.

An interesting discussion on the space of harmonic spinors underlines the analogy to the separation of variables method \cite{3}. The eigenvector of the operator $P_1$ on $S^1$ and the eigenvector of $P_2$ on $S^2$ for $P=P_1 +P_2$ represent the specific example of harmonic spinors.

In physics, spin structure arises due to the factorization of the Klein-Gordon (KG) operator to first order linear differential operators. It leads to the Dirac equation. In a Riemannian space the spin structure is an intricate issue. A connection akin to Levi-Civita connection and covariant derivatives on spin space  could be defined \cite{5}. The expressions for the spinor Laplacian and the KG equation for spinor are given below.
\begin{equation}
\Delta_{spin} \equiv -\frac{1}{4} (\gamma^\mu \gamma^\nu +\gamma^\nu \gamma^\mu) \nabla_\mu \nabla_\nu -\frac{1}{2} (\gamma^\mu \gamma^\nu) (\nabla_\mu \nabla_\nu -\nabla_\nu \nabla_\mu) \equiv -\nabla^\mu \nabla_\mu+ \frac{1}{4} R
\end{equation}
\begin{equation}
(-\nabla^\mu \nabla_\mu+ \frac{1}{4} R +m^2) \Psi =0
\end{equation}
Here $R$ is the scalar curvature. The notation in \cite{5} for covariant derivative and spin connection could be refined and made more transparent following the Infeld-van der Waerden spinor analysis \cite{9}. Here, our interest is on the spinors in flat space-time, evidently. $R=0$, and usual notation is applicable. Eq.(19) reduces to the standard Dirac equation for 4-spinor. 

Instead of abstract mathematical problem on harmonic spinors \cite{3,4} we develop physics-oriented approach to get new results on harmonic spinors in terms of vortex structures. The first important point is that the Dirac 4-spinor satisfying the Dirac equation also satisfies the KG equation (20) with $R=0$, therefore, any solution of the Dirac equation is a harmonic spinor by definition. The widely assumed plane wave solution
\begin{equation}
\Psi = \Psi_0 e^{i({\bf k} \cdot {\bf r} -\omega t)}
\end{equation}
is a harmonic spinor having trivial topology. Here ${\bf k} =\frac{{\bf p}}{\hbar}; ~ \omega = E/\hbar$. It is often convenient to use natural units $\hbar =c =1$. This case is similar to the trivial harmonic vector field (2). Since nontrivial topology is revealed by harmonic vector field of the form (1), we investigate harmonic spinors possessing nontrivial structure in the same spirit. The plane wave solution (21), in fact, has to be written in terms of creation and annihilation operators in the second quantized field theory. However, for the present discussion it is not necessary, therefore, only for the sake of completeness it is given in the Appendix.

Let us consider massless spinor field, i. e. specializing to the massless case of the Dirac equation. For the massless case, Eq.(19) for $R=0$ splits into Weyl equations for two 2-spinors
\begin{equation}
\Psi =\begin{bmatrix} \Psi_L \\ \Psi_R \end{bmatrix}
\end{equation}
In the Weyl equation the role of space reflection symmetry becomes important. Using chiral projections
\begin{equation}
P_L = \begin{bmatrix} I_2 & 0 \\ 0 & 0 \end{bmatrix}, ~ P_R = \begin{bmatrix} 0 & 0 \\ 0 & I_2 \end{bmatrix}
\end{equation}
\begin{equation}
P_L = \frac{I-\gamma^5}{2}; ~P_R =\frac{I+\gamma^5}{2}
\end{equation}
The $\gamma^5$ matrix in the chiral representation is
\begin{equation}
\gamma^5 = \begin{bmatrix} -I_2 &0 \\ 0 & I_2 \end{bmatrix}
\end{equation}
Here $I, ~I_2$ are $4 \times 4, ~2\times 2$ unit matrices respectively. The Weyl equations are given by
\begin{equation} 
{\bf \sigma} \cdot {\bf \nabla}  \Psi_L = \frac{\partial \Psi_L}{\partial t}
\end{equation}
\begin{equation} 
{\bf \sigma} \cdot {\bf \nabla}  \Psi_R =- \frac{\partial \Psi_R}{\partial t}
\end{equation}

Starting from the Weyl equations (26)-(27) it is straightforward and obvious that the Weyl spinors satisfy the wave equation with the operator $(\nabla^2 -\frac{\partial^2}{\partial t^2})$ making use of the identity $({\bf \sigma}\cdot {\bf \nabla})({\bf \sigma}\cdot {\bf \nabla})=\nabla^2~ I_2$. It is interesting to note that for the Pauli spinor assuming the Hamiltonian 
\begin{equation}
H_P ={\bf \sigma} \cdot {\bf p}
\end{equation}
multiplying the Schr$\ddot{o}$dinger equation
\begin{equation}
H_P \Psi_P =E \Psi_P
\end{equation}
by ${\bf \sigma} \cdot {\bf p}$ on both sides and using Eq.(29)  gives
\begin{equation}
-\nabla^2 \Psi_P =E^2 \Psi_P
\end{equation}
Therefore, the solution of the wave equation for the Weyl spinor or Eq.(30) for the Pauli spinor with a characteristic vortex structure could be obtained reducing them to the equation with the transverse operator
\begin{equation}
\nabla_t^2 = \frac{1}{r} \frac{\partial}{\partial r}\frac{r~ \partial}{~\partial r} +\frac{1}{r^2} \frac{\partial^2}{\partial \theta^2}
\end{equation}
Here the transverse Laplacian operator is written in the cylindrical coordinate system $(r, \theta, z)$.

Without loss of generality, it is sufficient to discuss Eq.(27) for Weyl spinor and the wave equation satisfied by $\Psi_R$. The well-known separation of variables in cylindrical coordinates can be used assuming $(z-, t-)$ dependence of the form $e^{-i\omega t+ i k z}$ for the solution of the wave equation, however, the Weyl equation (27) constrains this dependence for the two components of the spinor $\Psi_R$ as 
\begin{equation}
[\sigma_1 \frac{\partial}{\partial x}+\sigma_2 \frac{\partial}{\partial y}+\sigma_3 \frac{\partial}{\partial z}]\begin{bmatrix} \xi \\ \eta \end{bmatrix}=-I_2 \frac{\partial}{\partial t}\begin{bmatrix} \xi \\ \eta \end{bmatrix}
\end{equation}
In view of Eq.(32) we assume two sets of solutions
\begin{equation}
\begin{bmatrix} \xi_1 \\ \eta_1 \end{bmatrix}= \begin{bmatrix} \xi_0 f(r,\theta) e^{-i\omega t+ i k z} \\ \eta_0 f(r,\theta) e^{-i\omega t- i k z}\end{bmatrix}
\end{equation}
\begin{equation}
\begin{bmatrix} \xi_2 \\ \eta_2 \end{bmatrix}= \begin{bmatrix} \xi_0 f(r,\theta) e^{-i\omega t+ i k z} \\ \eta_0 f(r,\theta) e^{i\omega t+ i k z}\end{bmatrix}
\end{equation}
The wave equation in either case reduces to
\begin{equation}
\nabla_t^2 f(r,\theta) =0
\end{equation}
One of the solutions to this equation is given by
\begin{equation}
f(r, \theta) = r^l ~e^{-i l \theta}
\end{equation}

Substituting expression (36) in Eqs.(33) and (34) we get 2-spinor phase vortex: at $r=0$ the amplitude is zero but the phase is undetermined. Thus, we arrive at nontrivial topological harmonic spinors. Physical interpretation is possible in analogy with the definition of velocity field for nonrelativistic Schr$\ddot{o}$dinger wave equation. The probability current density can be used to define the velocity of the probability flow for the Schr$\ddot{o}$dinger wavefunction $\Psi_s$
\begin{equation}
{\bf v}_s = \frac{\hbar}{2 i m} (\Psi_s^* {\bf \nabla} \Psi_s - ({\bf \nabla }\Psi_s^*)\Psi_s)
\end{equation}
For the relativistic Dirac equation Gordon decomposition can be used to define the velocity field, however the massless limit for Gordon decomposition is not possible \cite{10}.  A detailed discussion on this issue is given in \cite{11}. Here we assume the following definition for the 2-spinor velocity field 
\begin{equation}
{\bf v} = \frac{\Psi_R^\dagger {\bf \nabla} \Psi_R - {\bf \nabla} \Psi_R^\dagger \Psi_R}{2 i \Psi_R^\dagger \Psi_R}
\end{equation}
It is important to note that the term velocity field for ${\bf v}_s$ is legitimate in view of the presence of the ratio $\hbar /m$ in the definition. However, in the definition (38) the term velocity field is used in analogy though it has the dimension of $(length)^{-1}$. Therefore, in contrast to vorticity $ {\bf \nabla} \times {\bf v}_s$ for the  Schr$\ddot{o}$dinger wavefunction we can interpret $\hbar ~({\bf r} \times {\bf v})$ as angular momentum for the spinor field.

The velocity fields for the two cases (33) and (34) calculated using the definition (38) are given by
\begin{equation}
{\bf v}_1 = \frac{l}{r} \hat{e}_\theta
\end{equation}
\begin{equation}
{\bf v}_2 = \frac{l}{r} \hat{e}_\theta +k \hat{e}_z
\end{equation}
Here $(\hat{e}_r, \hat{e}_\theta, \hat{e}_z)$ are unit vectors in cylindrical system. The azimuthal velocity component $\frac{l}{r} \hat{e}_\theta$ for both cases represents a singular vortex with topological defect. For the second case, the longitudinal velocity in z-direction corresponds to a propagating vortex. Since the radial velocity component is zero, the velocity vector field written in the cartesian coordinate system is
\begin{equation}
{\bf v}_{vortex} = -\frac{l y}{x^2 +y^2} \hat{i} +\frac{l x}{x^2 +y^2} \hat{j}
\end{equation}
It can be recognized that this velocity field is exactly of the form of the harmonic vector field (1), i. e ${\bf A}_h$. The vorticity is $\Omega={\bf \nabla} \times {\bf v}_{vortex}$, and its closed loop integral enclosing the origin gives quantized vorticity.

Irrespective of the physical interpretation of the vortex fluid flow, the present analysis leads to new mathematical results on the harmonic spinors: (1) spinors defined by partial differential operators could be the harmonic spinors of two types, namely trivial without topological defects, and having nontrivial singular or topological structures, for example, phase singularities or phase vortices. And, (2) there could exist a nontrivial harmonic vector field associated with nontrivial harmonic spinor, for example, ${\bf v}_{vortex}$ associated with Weyl 2-spinor. The first result is implicit in \cite{3} as we discuss in the next section while the second result has a potential to offer new approach to the subject of harmonic spinors and topology.

We can also calculate a vector quantity $\frac{\Psi_R^\dagger {\bf \sigma} \Psi_R}{\Psi_R^\dagger \Psi_R}$ for the two cases obtaining 
\begin{equation}
{\bf S}_1 = \frac{1}{\xi_0^2 +\eta_0^2} (2 \xi_0 \eta_0 (\cos 2kz~ \hat{i} -\sin 2kz ~\hat{j}) -(\xi_0^2 -\eta_0^2)~ \hat{k})
\end{equation}
\begin{equation}
{\bf S}_2 = \frac{1}{\xi_0^2 +\eta_0^2} (2 \xi_0 \eta_0 (\cos 2\omega t~ \hat{i} +\sin 2 \omega t ~\hat{j}) +(\xi_0^2 -\eta_0^2)~ \hat{k})
\end{equation}
Usually one expects these pseudo-vector quantities to be related with spin. However, we do not pursue this issue further and restrict to few remarks only. The first two terms in ${\bf S}_1$ for the special case of $\xi_0 = \eta_0$ survive and show spatial oscillatory behavior, while
in ${\bf S}_2$ we have oscillations in time. Note that in the extended particle picture this would correspond to the internal oscillatory behavior. 

\section{\bf ${\bf Z}_2$ Vortex}

Rather than the solution of the partial differential equations one could consider spinors as eigenfunctions of algebraic matrices; in nonrelativistic quantum mechanics a complex Hermitian Hamiltonian ${\bf r} \cdot {\bf \sigma}$ has been used in connection with the geometric phases. A real Hermitian Hamiltonian considered in \cite{7} has interesting topological implications. Wilczek \cite{7} assumes
\begin{equation}
H_W = \begin{bmatrix} x & y \\ y &-x \end{bmatrix}
\end{equation}
The Hamiltonian (44) is just $y \sigma_1 + x \sigma_3$. The solution of the Schr$\ddot{o}$dinger equation $H_W= E ~\Psi_W$ gives the energy eigenvalues $E_\pm = \pm \sqrt{x^2 + y^2}$, and the positive energy eigenfunction is
\begin{equation}
\Psi_+ = \begin{bmatrix} \cos \frac{\theta}{2} \\ \sin \frac{\theta}{2} \end{bmatrix}
\end{equation}
The crossing of the energy levels occurs as $\theta$ changes from $0~ \rightarrow ~2 \pi$. The energy level crossing is accompanied with the reversal of the sign of $\Psi_+$. This sign change is a discrete topological property; it is termed a $Z_2$ vortex. 

Instead of a real wavefunction (45) if a complex one
\begin{equation}
\Psi_+^\prime =e^{i \frac{\theta}{2}} \begin{bmatrix} \cos \frac{\theta}{2} \\ \sin \frac{\theta}{2} \end{bmatrix}
\end{equation}
is defined then the nontrivial topology seems to disappear: the wavefunction (46) is invariant under the transformation $\theta ~\rightarrow ~ \theta +2 \pi$. A deeper insight is gained using the Aharonov-Bohm $U(1)$ phase factor for the complex wavefunction $\Psi_+^\prime$. The Aharonov-Bohm like pure gauge potential
\begin{equation}
A_\theta = -\frac{i}{2}
\end{equation}
is shown to establish the equivalence of real and complex descriptions.

Relating this example with the theory of harmonic spinors leads to interesting results on the relationship between the continuous Lie groups and discrete groups. First we note Wilczek's observations. Let us divide the circle in two patches $-\pi -\epsilon_0 ~< \theta ~< \epsilon_0$ and $ -\epsilon_0 ~< \theta ~< \pi +\epsilon_0$, with $\epsilon_0$ an infinitesimally small positive number. The real wavefunction (45) on the two patches differ by a sign around $\theta=\pm \pi$. The suggested interpretation \cite{7} is that either one has a global basis for the wavefunction along with a nontrivial Aharonov-Bohm U(1) phase factor or smooth real-valued basis defined in two patches on the circle with a nontrivial transition $Z_2$.

In the context of harmonic spinors \cite{3} if $H$ is a finite dimensional space of harmonic spinors on $S^1$ then there are two spin structures: trivial line bundle with dimension of $H$ equal to 1, and Hopf bundle with  dimension equal to 0. Note that $H^1(S^1, Z_2) \equiv Z_2$. This topology has ramifications in Wilczek's example as well as the trivial and nontrivial harmonic vector fields discussed earlier in the present paper.  

A physical interpretation of the gauge potential (47) in terms of the intrinsic spin angular momentum has been put forward in \cite{12}. In fact, angular momentum holonomy as a physical mechanism for the geometric phases proposed earlier \cite{13} served the basis for this interpretation of pure guage potential (47) that appears in the geometric matrix in \cite{7}. In addition, the role of the matrix
\begin{equation}
C= \begin{bmatrix} 0 & 1 \\ -1 &0 \end{bmatrix}
\end{equation}
equivalent to the imaginary unit $i$, namely, $C^2 =-I_2$ was discussed to underline the difference between the group spaces for $U(1) $ and $SO(2)$. 

Here we seek new insights based on the characteristics of nontrivial harmonic spinors discussed in the preceding section. First, we calculate 
 the vector field defined by Eq.(38) for the real wavefunction (45)
\begin{equation}
{\bf v}_+ = [\Psi_+^\dagger (\frac{1}{r} \frac{\partial }{\partial \theta} ) \Psi_+ - ( (\frac{1}{r} \frac{\partial }{\partial \theta} ) \Psi_+^\dagger) \Psi_+] /2\pi i ~\hat{e}_\theta=0
\end{equation}
Next, we calculate this vector field for the complex wavefunction (46)
\begin{equation}
{\bf v}^\prime_+ = [\Psi_+^{\prime \dagger} (\frac{1}{r} \frac{\partial }{\partial \theta} ) \Psi_+^\prime - ( (\frac{1}{r} \frac{\partial }{\partial \theta} ) \Psi_+^{\prime \dagger}) \Psi_+^\prime] /2\pi i ~\hat{e}_\theta=\frac{1}{2 r} ~\hat{e}_\theta
\end{equation}

In the first case, that of the real wavefunction the velocity field vanishes, while for the complex wavefunction the velocity field is exactly the same as the Aharonov-Bohm gauge potential (47) calculated in a different way \cite{7}. It cannot be accidental; this is a remarkable result. The negative sign and the presence of imaginary unit $i$ in the gauge potential are due to the definition of the geometric phase factor used in \cite{11}
\begin{equation}
A_\theta=-\Psi_+^\dagger e^{-i\frac{\theta}{2}} \frac{\partial}{\partial \theta} e^{i\frac{\theta}{2}} \Psi_+
\end{equation}
We arrive at a third interpretation: nontrivial topology of real wavefunction (45) has zero velocity field (a trivial case), while the smooth global wavefunction (46) is associated with a velocity field having nontrivial harmonic structure of Eq.(1) as $(\hat{e}_\theta =-\sin \theta ~\hat{i} +\cos \theta ~\hat{j})$. The tentative physical interpretation \cite{12} of (47) in terms of spin finds support using the harmonic spinors and vector fields in view of the expression (50) for the velocity field. 

\section{\bf Continuous and Discrete Groups}

The Cartan classification of the Lie algebra of the classical continuous symmetry groups, for example, $SU(M)$, and Berger's classification of the holonomy groups are well-known. Salamon \cite{4} has listed holonomy groups for the Ricci-flat spin manifolds based on parallel spinors. The present study suggests a mathematically natural question: could the nontrivial harmonic spinors and the associated vector fields throw light on the classification of the continuous symmetry groups in conjunction with the discrete groups ? The role of $Z_2$ in the spin structure of $SO(2) \equiv U(1)$ is noted in \cite{3}. For the Lorentz group $SO(3,1)$ Weinberg \cite{14} points out the significance of the group $SL(2,C)/Z_2$. The group $SL(2,C)$ is the group of $2\times 2$ complex matrices with determinant equal to 1. The topology of this group is that of $R^3 \times S^3$, and it is simply-connected. On the other hand, the topology of $SL(2,C)/Z_2 \equiv R^3 \times S^3/Z_2$ is doubly-connected as $S^3/Z_2$ is three dimensional spherical surface with the opposition points on the sphere identified. 

Let us examine $SU(2)$ group and its Lie algebra in this perspective. It is known that the covering group of the rotation group $SO(3)$ is $SU(2)$, that is, there is a  two-to-one homomorphism $SU(2) ~\rightarrow ~ SO(3)$. The closed Lie algebra $su(2)$ is generated by the sigma matrices $\sigma_i, ~i=1,2,3$
\begin{equation}
[\sigma_i,~ \sigma_j] =2 i~ \epsilon_{ijk} ~\sigma_k
\end{equation}
Among the three sigma matrices, only one, namely $\sigma_2$ is complex
\begin{equation}
\sigma_2 = \begin{bmatrix} 0 & -i \\ i & 0 \end{bmatrix}
\end{equation}

If we recall Cartan's geometric interpretation of the spinors \cite{15} then complex spinor is in some sense a directed or polarized isotropic vector. In 3D Euclidean space a zero-length vector defines a pair of complex quantities called spinors and a vector ${\bf X}$ has associated with it a matrix
\begin{equation}
X = \begin{bmatrix} x_3 & x_1 - i x_2 \\ x_1+ i x_2  &  -x_3 \end{bmatrix}
\end{equation}
The sigma matrices are associated with the basis vectors in Eq.(54)  as can be seen writing $X = {\bf X} \cdot {\bf \sigma}$. In the pseudo-Euclidean space in 3D the matrix associated with ${\bf X}$ is real
 \begin{equation}
X = \begin{bmatrix} x_3 & x_1 -  x_2 \\ x_1+  x_2  &  -x_3 \end{bmatrix}
\end{equation} 
The basis matrices for (55) are also real $\sigma_1,~\sigma_3,~ C^T$. Here $C^T$ is transpose of C-matrix (48). 

In analogy to the role of $Z_2$ in the Abelian group  $U(1)$ discussed in the preceding section and also in \cite{12} we propose that the discrete symmetry group $SL(2, Z)$ has a similar role in connection with $SU(2)$. The generators of the non-Abelian group $SL(2,Z)$ are $C^T$ and T
\begin{equation}
T = \begin{bmatrix} 1 & 1 \\ 0 & 1 \end{bmatrix}
\end{equation}
Note that $SL(2, R)$ is the group of $2\times 2$ real matrices with determinant equal to 1. Just like the integers within real numbers, the group $SL(2, Z)$ is a discrete group in $SL(2,R)$. It would be interesting to give a mathematical proof of this proposition and explore whether such kind of relationship between the continuous Lie groups and discrete groups holds in general.  

\section{ \bf Conclusion}

The present paper is limited to phase vortices for 2-spinors, however, in view of numerous experiments on the vortices in high energy electron beams since the notable work of Uchida and Tonomura \cite{16}, it would be of great value to extend this approach to Dirac 4-spinors. The present approach on harmonic spinors offers a possible new perspective on the physics of quantized and half-quantized vortices in terms of topology of the spinors. 

To conclude, a new physics-oriented approach to gain new insights on the harmonic spinors with nontrivial topology is developed in the present work.

{\bf APPENDIX}

The plane wave solution of the Dirac equation in the second quantization is represented by
\begin{equation}
\Psi ({\bf r},t) = \int \frac{d^3{\bf p}}{2 (2\pi)^3 E_p} \sum_{s=1}^2 [u({\bf p},s)  a({\bf p},s) e^{-i({\bf p}\cdot {\bf r} -E t)} + v({\bf p},s)  b^\dagger ({\bf p},s) e^{+i({\bf p}\cdot {\bf r} -E t)}
\end{equation}
Here $a^\dagger,~a$ are the creation and annihilation operators for particle (electron) states respectively satisfying the anti-commutation rules; $b^\dagger,~b$ for the anti-particle states.

CONFLICT OF INTEREST:~ The author declares no conflicts of interest.

\end{document}